\documentclass[11pt]{article}
\usepackage{moriond,epsfig}

\def\Journal#1#2#3#4{{#1} {\bf #2}, #3 (#4)}

\def\NIMA{{\em Nucl. Instrum. Methods} A}
\def\NPB{{\em Nucl. Phys.} B}
\def\PLB{{\em Phys. Lett.}  B}
\def\PRL{\em Phys. Rev. Lett.}
\def\PRD{{\em Phys. Rev.} D}

\def\sst{\scriptscriptstyle}

\def\be{\begin{equation}}
\def\ee{\end{equation}}
\def\bea{\begin{eqnarray}}
\def\eea{\end{eqnarray}}

\begin{document}
\vspace*{4cm}
\title{QCD CHALLENGES AT THE LHC}

\author{ VITTORIO DEL DUCA }

\address{Istituto Nazionale di Fisica Nucleare, Sez. di Torino\\
via P. Giuria, 1 - 10125 Torino, Italy}

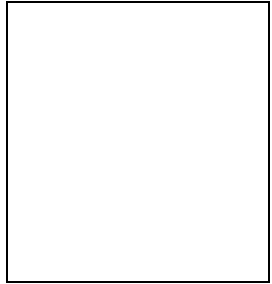
\maketitle\abstracts{
In this talk I review some challenges which await perturbative QCD at the 
Large Hadron Collider. In particular, I consider the underlying event,
Monte Carlo methods and next-to-leading order (NLO) calculations.}

\section{Introduction}\label{sec:intro}

With the start of LHC operations next year,
high-energy physics will enter a new era of discovery.
The LHC is a proton-proton collider that will
function at the highest energy ever attained in the laboratory,
and will probe a new realm of high-energy physics.
The use of a high-energy hadron collider as a research tool makes 
substantial demands  upon the  theoretical 
understanding  and the predictive  power  of  QCD,
the  theory  of  the strong  interactions within the Standard Model.
In the non-perturbative (low $Q^2$) regime several approaches to QCD,
like lattice gauge theory, Regge theory, chiral perturbation theory,
large $N_c$, heavy-quark or soft-collinear effective theories, are used.
In this talk I concentrate on the perturbative (high $Q^2$) regime, where
QCD purports to be a precision toolkit for exploring Higgs and
Beyond-the-Standard-Model (BSM) physics. Precision QCD aims to
achieve an ever more precise determination of the strong coupling
constant $\alpha_{\sst S}$, of the parton distribution functions (p.d.f.), of the
electroweak parameters, of the LHC parton luminosity, and of the
strong corrections to Higgs and BSM signals and to their backgrounds.

\section{Breaking factorisation}\label{sec:fact}

The tenets of perturbative QCD (pQCD) are the universality of
the infrared (IR) behaviour, the cancellation of the IR singularities
for suitably defined variables, like jets
and event shapes, and in the case of hadron-initiated processes, like
electron-proton or (anti)proton-proton collisions, the factorisation of
the short- and long-range interactions. 
Factorisation in proton-proton ($pp$) collisions states that the
cross section for the production of high-mass states, characterised by the
large scale $Q^2$, can be separated into a parton cross section
for the primary event, and into p.d.f.'s,
measurable experimentally and whose evolution is described by the DGLAP
equations.

Outside of the realm of factorisation lies the underlying event (UE),
which can be operatively defined as whatever is in the $pp$ interaction
besides the primary scattering. In particular, UE
includes the multiple-parton interactions as well as the interaction
of spectator partons, {\it i.e.} other than the ones initiating the primary 
scattering. The assumption is 
that if such an interaction occurs it is characterised by a scale 
$\Lambda$ of the order of a GeV, and so it is suppressed by powers of 
$\Lambda^2/Q^2$ with respect to the primary scattering.
Thus, UE breaks factorisation by means of power-suppressed
contributions. How important are they for a precision calculation ?
There is no obvious answer to this question since of course we cannot use
the pQCD framework to model UE, and its analysis must rely solely upon 
the data. In $p\overline{p}$ collisions at the Tevatron, UE is being 
studied~\cite{Acosta:2004wq} by analysing in single-jet production
the charged-particle multiplicity in regions 
which are perpendicular in azimuth to the jet, since that region
is expected to be sensitive to UE. A modelling of the data is then
performed through the shower Monte Carlo (MC) PYTHIA. The UE sensitivity to
beam remnants and to multiple interactions can be reduced by selecting
back-to-back two-jet topologies. A similar investigation is 
being planned also through the Drell-Yan production of vector bosons.
Understanding and modelling UE at LHC will represent a major challenge.

Other examples of factorisation-breaking contributions are $a)$ the power
corrections: MC and theory modelling of power corrections were 
laid out and tested at LEP, where they provided an accurate determination of 
$\alpha_S$~\cite{Dokshitzer:1995qm}. However, models still need be tested in 
hadron collisions: a study of single-jet production at Tevatron running at
two different centre-of-mass energies shows that the Bjorken scaling is violated
more than logarithmically, and data can fitted by assuming a power-correction
shift in the jet $E_T$~\cite{santabarb}; $b)$ diffractive 
events~\cite{Collins:1992cv}, which are known to violate factorisation
at Tevatron~\cite{Terashi:2000vs,Acosta:2003xi}.

\section{Monte Carlo models}\label{sec:mc}

The detection of Higgs and BSM signals requires a precise modelling of
their backgrounds. Examples are QCD production of $W + 4$ jets and
of $W W + 2$ jets, which are backgrounds to Higgs production through
vector-boson fusion (VBF) with the Higgs decaying into a $W W$ pair, as well
to $t\overline{t}$ production, or $W + 6$ jets and $W W + 4$ jets,
which are backgrounds to $H t\overline{t}$ production. One approach is to model 
QCD production through matrix-element MC
generators, which provide an automatic computer generation of
processes with many jets, and/or vector or Higgs bosons.
There are several such multi-purpose generators, 
like e.g. ALPGEN~\cite{Mangano:2002ea},
MADGRAPH/MADEVENT~\cite{Stelzer:1994ta,Maltoni:2002qb}, 
COMPHEP~\cite{Pukhov:1999gg}, GRACE/GR@PPA~\cite{Ishikawa:1993qr,Yuasa:1999rg}, 
HELAC~\cite{Kanaki:2000ey}, and SHERPA~\cite{Krauss:2004bs} (which has got its 
own showering and hadronisation).
A different example is PHASE/PHANTOM~\cite{Accomando:2004my}, 
a MC generator dedicated to processes
with six final-state partons only, thus suitable to $t\bar{t}$ production,
$W W$ scattering, Higgs production via VBF and 
vector-boson gauge coupling studies, but where no approximation is used.
Matrix-element MC generators are particularly suitable to studies which
involve the geometry of the event, because the jets in the final state are
generated at the matrix-element level, and thus exactly at any angle.
In addition, they can be interfaced to parton-shower MC generators, 
like HERWIG~\cite{Marchesini:1991ch} or PYTHIA~\cite{Sjostrand:1993yb}, to
include showering and hadronisation. Furthermore, a procedure 
(CKKW)~\cite{Catani:2001cc,Krauss:2002up} has been devised to interface parton
subprocesses with a different number of final states to parton showers.
Finally, MC@NLO~\cite{Frixione:2002ik}: a procedure and a code to match exact 
NLO computations to shower MC generators. In a way, this is the most desirable
procedure, because it embodies the precision of NLO partonic calculations
in predicting the overall normalisation of the event,
while generating a realistic event set up through showering and hadronisation.
It cannot be, though, multi purpose, being obviously limited to the 
processes for which the NLO corrections are known. Challenges in this instance
are to include as many NLO processes as possible for Higgs and BSM signals and
for their backgrounds, as well as to extend the CKKW approach to it.

\section{NLO calculations with many jets}\label{sec:nlo}

NLO calculations have several desirable features. $a)$ the jet structure:
while in a leading-order
calculation the jets have a trivial structure because each parton becomes
a jet, to NLO the final-state collinear radiation allows up to two partons to
enter a jet; $b)$ a more refined p.d.f. evolution through the initial-state
collinear radiation; $c)$ the opening of new channels, through the inclusion
of parton sub-processes which are not allowed to leading order; $d)$ a
reduced sensitivity to the (fictitious) renormalisation and factorisation 
scales allows to predict the normalisation of physical observables, which
is usually not accurate to leading order. That is the first step toward
precision measurements in general, and in particular toward an accurate
estimate of signal and background for Higgs and New Physics at LHC;
$e)$ finally, the matching with MC@NLO mentioned in Sect.~\ref{sec:mc}.

IR singularities appear in the intermediate
stages of a NLO calculation. However, the structure of QCD is such that those
singularities are universal, {\it i.e.} they do not depend on the
process under consideration, but only on the partons involved in
generating the singularity. Thus, in the 90's process-independent
procedures~\cite{Giele:1991vf,Giele:1993dj,Frixione:1995ms,Nagy:1996bz,Catani:1996vz}
were devised to regulate those divergences, which use universal 
counterterms to subtract the divergences. However, a look at the 
history of NLO calculations shows that given a certain process, it is often 
very time-consuming to compute to NLO the 
addition of even just one more jet to it. 
Why in a NLO calculation is it so difficult to add more 
particles in the final state ? The loop integrals occurring in the
virtual contributions are involved and process dependent.
In addition, more final-state particles imply more scales in the process,
and so lenghtier analytic expressions in the loop integral. 
In fact, except for special cases like the NLO corrections to the
electroweak production of a vector-boson pair $+$ 2 jets~\cite{Jager:2006zc},
there are no processes with more than three final-state particles for which
the NLO corrections are known.
Recently, a twistor-inspired approach~\cite{Witten:2003nn}, which has
allowed for great advances in the analytic computation of tree and one-loop
amplitudes~\cite{Cachazo:2004kj,Britto:2005fq,Bern:2005hs}, 
as well as several semi-numerical approaches which show promise to handle NLO 
corrections in an automated 
way~\cite{Kramer:2002cd,Binoth:2002xh,Nagy:2003qn,Giele:2004iy,Binoth:2005ff,Ellis:2005qe,Anastasiou:2005cb,Czakon:2005rk,Binoth:2006rc}, 
have surfaced. However, the programme
of applying sistematically NLO computations to studies of signals and
backgrounds for Higgs and New Physics represents yet a major challenge,
which will be undoubtedly receiving much attention in the next future.

\section*{Acknowledgments}
Work supported by MIUR under contract 2004021808--009.

\end{document}